\documentstyle[prl,epsf,aps]{revtex}
\def\c60{$\rm C_{60}$}
\def\nec60{Ne@\c60}
\def\hec60{He@\c60}
\def\etal{{\it et al.}}

\begin{document}
\draft

\preprint{to appear in Phys.\ Rev.\ Lett.\ {\bf 80} (1998)}
\twocolumn[\hsize\textwidth\columnwidth\hsize\csname @twocolumnfalse\endcsname
\title{Theory of Spontaneous Polarization 
of Endohedral Fullerenes}

\author{D.\ P.\ Clougherty and F.\ G.\ Anderson} 

\address{Department of Physics, University of Vermont, Burlington, VT 
05405-0125}

\date{August 25, 1997}

\maketitle

\begin{abstract}
A pseudo--Jahn--Teller model describing central atom
distortions is proposed for endohedral 
fullerenes of the form A@\c60 where A is either a rare gas 
or a metal atom. A critical (dimensionless) coupling $g_c$ is found, below which the
symmetric configuration is stable and above which inversion symmetry is broken. Vibronic
parameters are given for selected endohedral fullerenes.

\end{abstract} 

\vskip2pc]

\narrowtext

Shortly after it was established that fullerenes are closed-cage
structures of carbon, investigators speculated that atoms and
clusters might be coaxed to occupy the space inside the cage. The
existence of such endohedral fullerenes is now well-established by
experiment\cite{bethune}. 
Such endohedral doping of \c60 shows some promise as a technique to
modify fullerene properties such as electric polarizibility,
optical and infrared absorption, and magnetic susceptibility. 

Several numerical studies\cite{nec60,liu,dunlap,wang,li} have concluded that
for the case of metal endohedral dopants, such as Na@\c60, Li@\c60, and
Ca@\c60, the metal atom transfers its valence electrons to the \c60
cage, and establishes an off-center equilibrium position, resulting in a net
electric dipole moment for the molecule. For the case of rare gas dopants,
however, the central atom sits at the cage center. 

A classical model was used
by Erwin\cite{erwin} to explain the two disparate cases; for metal dopants, the
charge transferred to the cage occupies a highly delocalized t$_{1u}$
orbital. Erwin treats the resulting negatively charged cage as a perfectly
conducting sphere. The total energy of the metal cation and the conducting
sphere is minimized by placing the cation off-center where electrostatic
 interaction between the cation and the cage can be made more
attractive as the charge on the cage is polarized. The distortion is
stabilized in Erwin's model by including a 
short-ranged repulsive interaction of Lennard-Jones form for the
electron overlap between 
the cation and cage electrons.

Such a classical model would predict that for the case of metal dopants,
inversion symmetry is broken as the central atom is stabilized in an
off-center equilibrium position; while for closed-shell 
rare gas atoms, there will be no charge transfer and, consequently, no central
atom distortion. However, this classical model suffers from two serious
problems. First, for the case of singly-ionized alkali atoms, such a
classical model would again predict no charge transfer, as such an atom is
isoelectronic to a rare gas atom. Thus, charged metallofullerenes such as
Li$^+$@\c60 would have the cation symmetrically positioned according to the
classical model. This, however, contradicts electronic structure
calculations\cite{nec60,liu}. Second, for the distortive case of
metallofullerenes, the net dipole moment calculated from the classical model is
zero, as the contributions from the cation and the polarized cage exactly
cancel.

The purpose of this work is two-fold: (1) to remedy
the shortcomings of the classical model, and (2) to give insight into the 
physical mechanisms underlying previous numerical studies.
 We present
here a model for the distortion mechanism where the electronic states are 
treated quantum mechanically. The propensity of the
central atom for distortion can be understood as arising from the 
pseudo--Jahn--Teller (PJT) effect, where excited electronic states are mixed
with highest occupied molecular orbital (HOMO)
through a symmetry-breaking distortion.
We note that PJT models have been successfully invoked 
in other contexts to 
describe, for example, 
distortions in the case of the F-center\cite{ham} and in oxygen and
nitrogen impurities in silicon\cite{anderson}.

We give analytical expressions
for the magnitude of the distortion in this PJT model, and we find that
there is a critical 
vibronic coupling
$g_c$; for values of
$g > g_c$, the dopant moves off-center, and for $g\le g_c$, the dopant
sits at the cage 
center. We also find that for linear coupling the adiabatic potential
surface is spherically 
symmetric, in good agreement with the numerical
calculations\cite{nec60,dunlap}, 
while the symmetry of the perturbative warping of the adiabatic
potential can be 
explained with the addition of higher order terms.

We detail a model for the PJT coupling 
that involves a displacement of the central ion in endohedrally doped 
$\rm{C}_{60}$.  
We start by describing the electronic structure of the 
$\rm{C}_{60}$ molecule.  The valence electrons that are of interest 
are the radially-directed $p$ 
electrons on the carbon atoms that take part in $\pi$ bonding.
There is one such electron on each carbon atom, the remaining valence electrons
taking part in $\sigma$ bonding and having a much lower energy.  From the
atomic $p$ orbitals on each carbon, we construct molecular orbitals
consistent with the $I_{h}$ symmetry of the $\rm{C}_{60}$ molecule.  However,
since $\rm{C}_{60}$ is nearly spherical\cite{dpc}, it is 
possible to associate the different manifolds of molecular orbitals with 
angular momentum states $L$.  The $\rm{C}_{60}$ cage 
is electrically neutral; hence, the HOMO
is the filled
$h_{u}$ level coming from the $L = 5$ molecular orbitals\cite{haddon}.
The lowest unoccupied molecular orbital (LUMO) is a $t_{1u}$ manifold, also
coming from
$L = 5$. Slightly higher in energy however is a $t_{1g}$ manifold, from $L =
6$.  It is this excited state that plays a crucial role in our PJT coupling
model.

We now consider the Jahn-Teller (JT) interaction involving the displacement of 
the central ion from its on-center position.  For the case of dopants 
isoelectronic to
the rare gases, the HOMO remains the filled
$h_{u}$ manifold. Such a state is non-degenerate, and consequently there
is no pure JT interaction.  There is, however, a 
PJT interaction involving the excited $t_{1g}$ manifold.
The $t_{1g}$ manifold is the one of interest since the displacement modes of 
the central ion $\vec Q$ span the irreducible 
representation $t_{1u}$, and $t_{1g}$ is contained in the direct product 
$h_{u} \otimes  t_{1u}$.

The Hamiltonian is taken to be
\begin{equation}
{\cal H}  =  {\cal H}_{o} + {\cal H}_{elas} + {\cal H}_{JT}
\label{hamiltonian}
\end{equation}
${\cal H}_{o}$ is the electronic Hamiltonian and gives the  
electronic energy of the $h_{u}$ and $t_{1g}$ manifolds for the fully symmetric
(I$_h$) system
\begin{equation}
{\cal H}_{o}  =  0 \ \vert h_{u}\rangle\langle h_{u} \vert  +  \Delta \ \vert
t_{1g}\rangle\langle t_{1g} \vert, 
\label{el}
\end{equation}
where $\Delta$ is the energy difference between the $t_{1g}$ and $h_{u}$
manifolds, $\Delta  =  {\rm E}(t_{1g}) - {\rm E}(h_{u})$.
${\cal H}_{elas}$ is the elastic energy Hamiltonian for a classical
displacement  
of the central ion. Assuming the harmonic
approximation,
\begin{equation}
{\cal H}_{elas}  =  {k \over 2}\> (Q_{x}^{2} + Q_{y}^{2} + Q_{z}^{2}) \>
{\bf I},
\label{vib}
\end{equation}
where {\bf I} is the electronic identity operator. 
Actually, $\vec Q$ includes
$t_{1u}$-symmetric distortions of the \c60 cage as well as a displacement of 
the central ion. 
But since the cage is much
more massive than the dopant atoms considered (large cage mass
approximation),  the 
collective displacement $\vec Q$ is well approximated by dopant displacement 
alone. 
 Lastly, we have the
${\cal H}_{JT}$ that describes the PJT coupling between the $h_{u}$ and
$t_{1g}$ manifolds resulting from a displacement of the central ion from its
on-center position, $(h_u\oplus t_{1g})\otimes t_{1u}$ coupling.
These matrix elements are 
determined from the coupling coefficients given by Fowler and
Ceulemans (Table~5 of Ref.~\cite{fowler}) where the coordinate axes are 
chosen to be coincident with the three orthogonal $C_{2}$ rotation
axes. In our eight 
dimensional electronic space, we have

\begin{equation}
{\cal H}_{JT} = \bordermatrix{&&\cr
&0&h_{JT}\cr
&h_{JT}^\dagger&0},
\label{pjt}
\end{equation}
where
\begin{equation}
h_{JT} = \bordermatrix{& t_{1x}& t_{1y} & t_{1z}\cr  
h_{z^2} & -V\, Q_{x} & -V \, Q_{y} & 2 \> V \, Q_{z}\cr  
h_{x^2 - y^2}& \sqrt{3} \> V \, Q_{x}& -\sqrt{3} \> V \, Q_{y} & 0\cr  
h_{\xi} & 0 &  \sqrt{3} \> V \, Q_{z} & \sqrt{3} \> V \, Q_{y} \cr
h_{\eta} & \sqrt{3} \> V \, Q_{z}  &  0 & \sqrt{3} \> V \, Q_{x} \cr
h_{\zeta}& \sqrt{3} \> V \, Q_{y}  &  \sqrt{3} \> V \, Q_{x} & 0 }, 
\label{h} 
\end{equation}
and $V$ gives the strength of the PJT coupling.  We note that $h_{z^2} =
\sqrt{3/8} \; h_{\theta} - \sqrt{5/8} \; h_{\epsilon}$ and $h_{x^2 - y^2} = 
\sqrt{5/8} \; h_{\theta} + \sqrt{3/8} \; h_{\epsilon}$, in which $h_{\theta}$
and $h_{\epsilon}$ are the states employed in Ref.~\cite{fowler}.

We proceed by diagonalizing the Hamiltonian in Eq.~\ref{hamiltonian}.
By way of example, 
we consider the simple case in which $Q_{z} = Q$ and $Q_{x} = Q_{y} = 0$.
That is, we consider the particular case in which the central ion displaces 
along one of the $C_{2}$ symmetry axes.  In this instance, the electronic
Hamiltonian matrix reduces to three $2 \times 2$ blocks and two
 $1 \times 1$ blocks.
The $h_{x^2 - y^2}$ and the $h_{\zeta}$ states are unaffected by this 
particular distortion; and so, their energies remain unchanged.  
One $2 \times 2$ block, which mixes the states $h_{z^2}$ and $t_{1z}$,
 has the form
\begin{equation}
{\cal H}^{A} = \pmatrix{0  & c \> V \, Q \cr
 c \> V \, Q & \Delta}.\label{z2}
\end{equation}
where $c=2$.

The two remaining $2 \times 2$ blocks, mixing the pairs of states $\{ h_{\eta},
t_{1x} \}$ or  $\{ h_{\xi}, t_{1y} \}$, also have the form
given in Eq.~\ref{z2} with $c=\sqrt{3}$.

The energy shift for a distortion $Q$ is found by summing over the ten
lowest energy  
states, including spin degeneracy.  Hence, we find
\begin{equation}
{\rm E} = 2 \, \rm{E_{-}^{A}} + 4 \, \rm{E_{-}^{B}} +
  {{k \, Q^2} \over 2}, 
\label{energy}
\end{equation}
where $\rm{E_{-}^{A}}={\Delta \over 2} - \left[ ({\Delta \over 2})^{2} +
4 \, V^{2} \, Q^{2} \right] ^{1/2}$
and 
$\rm{E_{-}^{B}}={\Delta \over 2}- \left[ ({\Delta \over 2})^{2} +
3 \, V^{2} \, Q^{2} \right] ^{1/2}$.
We then minimize $\rm{E}$ with respect to 
$Q$.  We find that $Q_0 = 0$, corresponding
to no displacement of the central ion, gives the minimum energy configuration
when JT coupling is weak.  More precisely, if
\begin{equation}
g \equiv {V^{2} \over  k \, \Delta} \le {1\over 40},
\label{gc}
\end{equation}
there is no displacement of the central ion.  On the
other hand, if the JT coupling is strong enough, $g > \frac{1}{40}$,
the minimum  
energy configuration is found for $Q_0 \ne 0$.  That is, the total energy of
the endohedrally doped $\rm{C}_{60}$ molecule can be reduced if the 
central ion displaces from its on-center position.  The minimum energy
configuration is found when $\hat Q_0 \equiv V \,Q_0 / \Delta$ satisfies
the equation
\begin{equation}
{{2} \over {(1 + 16 \, \hat Q_0^{2})^{1/2}}} +
 {{3} \over {(1 + 12 \, \hat Q_0^{2})^{1/2}}} = {1 \over 8g}.
\label{distort}
\end{equation}

We now consider the most general displacement of the central ion.  Here,
we parametrize $Q_{x}$, $Q_{y}$, and $Q_{z}$ by $Q_{x} = Q \sin \theta \, 
\cos \phi$, $Q_{y} = Q \sin \theta \, \sin \phi$, and $Q_{z} = 
Q \cos \theta$.  We 
substitute this into the matrix for the full Hamiltonian.  The corresponding
characteristic equation, whose roots yield the energy eigenvalues, is 
found to be independent of $\theta$ and $\phi$.  Thus, the one-electron
energy eigenvalues are {independent of the direction of the displacement
of the central ion}, and the adiabatic 
energy surface is spherical. 

\begin{figure} [!t]
\centering
\leavevmode
\epsfxsize=7cm
\epsfysize=7cm
\epsfbox[18 144 592 718] {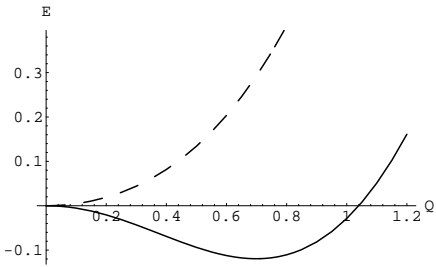}
\caption[]
{\label{v} Adiabatic potential $E$ vs. magnitude of t$_{1u}$ distortion Q for
Na$^+$@\c60. A second curve (dashed) with reduced vibronic coupling
$V$ ($g=0.02$) is shown 
where the symmetric configuration is stable.}
\end{figure}

Such an accidental SO(3) symmetry has been known to occur
in other Jahn-Teller systems with icosahedral symmetry \cite{pooler}.
The adiabatic potential $E$ as a function of
central atom displacement is plotted in Fig.~\ref{v} for $g > g_c$. When the
kinetic energy associated with the displacement $\vec Q$ is included in
Eq.~\ref{hamiltonian}, low frequency excitations corresponding to a 
rotation of
the distortion about the symmetric configuration are found, together
with higher frequency 
radial excitations. Such excitations have been previously discussed
elsewhere\cite{dunlap,li}. It is important to note 
that there is no geometric phase in this 
model. Thus, this model
establishes the quantization condition 
for the rotational wavefunctions
previously assumed\cite{dunlap,li}.

For the case of metallic dopants, the HOMO becomes the $t_{1u}$
manifold\cite{wastberg}. Here, such a partially filled state is subject to JT
distortions. In fact, it has been proposed that such coupling is responsible
for superconductivity in the alkali-doped fullerides\cite{super}. However,
the JT distortion is of $h_g$ symmetry, and thus does not contribute to a
dipole moment. 

In addition to the $t_{1u}\otimes h_g$ JT coupling, there is PJT coupling
between the $t_{1u}$ and $t_{1g}$ electronic states by a $t_{1u}$ distortion, 
$(t_{1u}\oplus t_{1g})\otimes t_{1u}$ coupling. 
We take the PJT coupling to be 
\begin{equation}
h_{JT} = \bordermatrix{& t_{1ux}& t_{1uy} & t_{1uz} \cr  
t_{1gx} & 0  &   V  Q_{z} & -  V  Q_{y} \cr
t_{1gy} & -  V  Q_{z} & 0 &   V  Q_{x} \cr  
t_{1gz} &    V  Q_{y} &  - V  Q_{x} & 0}.
\label{tham}
\end{equation}

Diagonalizing the electronic Hamiltonian yields the following
eigenvalues: 0, $\Delta'$, and
$
\rm{E_{\pm}} = {\Delta' \over 2} \pm \left[ ({\Delta' \over 2})^{2} +
 V^{2} \, Q^{2} \right] ^{1/2}$ where $\Delta'$ is the $t_{1u}$-$t_{1g}$ 
energy gap. 
Again, the
eigenvalues are invariant under rotations \cite{ceulemans}.

For alkali metal endohedral fullerenes with a $t^1_{1u}$ ground state, we find
that the energy shift for a distortion $Q$ is 
\begin{equation}
{\rm E}={\Delta' \over 2} - \left[ ({\Delta' \over 2})^{2} +
 V^{2} \, Q^{2} \right] ^{1/2} + {\frac{1}{2}{k \, Q^2}},
\label{teigen} 
\end{equation}
By minimizing the total energy, we conclude that there is no distortion
unless $g > \frac{1}{2}$.

A non-vanishing electric dipole moment results from the PJT
distortion. Because of SO(3) 
invariance,  we consider without loss of generality the case of a
simple axial distortion 
($Q_z \ne 0$) of an endohedral dopant isoelectronic to a rare gas
atom. 
We find that a
spontaneous dipole 
moment is formed for a non-vanishing distortion $Q_z$, in contrast to
the classical 
model of Erwin which assumes perfectly metallic screening. Using the vibronic
eigenstates of  Eq.~\ref{hamiltonian} together with the Clebsch-Gordan
coefficients obtained 
from Ref.~\cite{fowler}, we find an electric dipole moment in the
direction of the central 
atom distortion  of the form
\begin{equation}
p=8 p_0 \left({\alpha\over 1+\alpha^2}+{\sqrt{3}\beta\over 1+\beta^2}\right)
\label{dipole}
\end{equation}
where $p_0$ is a reduced matrix element, $\langle t_1 || p || h\rangle$,
$\alpha={\sqrt{1+16\hat Q_0^2}-1\over 4\hat Q_0}$, and $\beta={\sqrt{1+12\hat
Q_0^2}-1\over 2\sqrt{3}\hat Q_0}$.
That numerical
 calculations on light rare gas endohedrals find the
equilibrium of the rare gas 
dopant at the cage center indicates that $g$ for these systems is less than 
$\frac{1}{40}$. In the case of positively charged alkali endohedrals
such as Na$^+$@\c60 
and Li$^+$@\c60, previous calculations \cite{liu,li} have found an
off-center equilibrium 
position for the alkali ion, indicating that $g$ for these endohedrals exceeds
$\frac{1}{40}$.

Parameters for selected endohedrals are given in Table~\ref{params}.
For Na$^+$@\c60, we take $\Delta=2.8$ eV, $Q_0=0.7$\AA, and ${\rm E}=-0.12$ eV
\cite{dunlap}. From Eqs.~\ref{energy} and
\ref{distort}, we fit the results of Ref.~\cite{dunlap} with $V=0.69$ eV/\AA\ 
and $k=5.8$
eV/\AA$^2$, giving $g=0.03$. 

Assuming the 
primary coupling is between $t_{1u}$ and $t_{1g}$, 
we estimate the relevant parameters for Na@\c60 
on the basis of previous 
calculations. We take $\Delta'=1.0$ eV, $Q_0=1.5$\AA, and
${\rm E}=-1.4$ eV \cite{liu}. We fit the results with $V=2.47$ eV/\AA\ 
 and
$k=1.63$ eV/\AA$^2$, giving $g=3.73$.

Ca@\c60 can be treated within the same framework.
For Ca@\c60, the electronic ground state for I$_h$ symmetry is
$^3$T$_{1g}$. PJT 
coupling with the $^3$T$_{1u}$ state gives rise to a distortion that
breaks inversion 
symmetry. The calculations of Ref.~\cite{wang} can be fit with
$V=1.65$ eV/\AA\  
 and
$k=3.62$ eV/\AA$^2$, giving $g=0.58$.

The SO(3) invariance of the adiabatic potential within the linear PJT
coupling will be 
broken when higher order couplings are taken into account. The amount
of the warping 
of the adiabatic potential has been found \cite{nec60} to be
negligible when the 
central atom dopant is either a rare gas or a positively charged
alkali atom. Thus, the 
linear PJT model well describes the previous numerical
 calculations for these cases.

For the case of Ca@\c60, it has been found \cite{wang} that the
molecular symmetry is 
broken down to C$_{5v}$.
While the addition of
$^3$T$_{1g}\otimes$h$_g$ linear JT coupling for Ca@\c60 does not break SO(3)
invariance, D$_{5d}$ warping \cite{warping1} of the adiabatic
potential surface
does result from 
quadratic JT coupling
involving distortions of 
h$_g$ symmetry \cite{warping2}, 
leading to a molecular symmetry of C$_{5v}$. The
softest modes of 
\c60 are of h$_g$ symmetry; consequently, there is a small energy cost
for such 
distortions. 
For such a quadratic coupling to the low frequency h$_g$ vibrations, 
the directions of three-fold symmetry can contain relative maxima, while
the directions of 
two-fold symmetry contain saddle points. For weak coupling, such a
warping will be small 
compared to the PJT stabilization energy. 

We conclude that 
PJT coupling provides a unifying framework for
understanding the spontaneous polarization of endohedral fullerenes.
The parameters of 
our PJT model were obtained by fitting to previous numerical
calculations. We find that the 
structural distortion resulting from endohedral doping is yet another example
illustrating the importance of vibronic effects in shaping the
properties of fullerene 
systems.

We thank S.~Erwin for bringing his work to our attention.
Acknowledgment is made to the Donors of The Petroleum
Research Fund, administered by the American Chemical Society, for
support of this research. This research was supported in part by the National
Science Foundation under Grant No.\ PHY94-07194.

\begin{table}
\caption{Vibronic parameters for selected endohedrals.}
\begin{tabular}{lddddd}
Dopant&$\Delta$ or $\Delta'$ $(eV)$&$Q_0$ (\AA)&V (eV/\AA)&$k$ (eV/\AA$^2$)&$g$\\
\tableline
Na$^+$&2.8$^{\rm a}$&0.7$^{\rm d}$&0.69&5.78&0.03\\
Na&1.0$^{\rm a}$&1.5$^{\rm c}$&2.47&1.63&3.73\\
Ca&1.31$^{\rm b}$&0.83$^{\rm b}$&1.65&3.62&0.58\\
\end{tabular}
\label{params}
{$^{\rm a}$ Ref.~\cite{wastberg}}
{$^{\rm b}$ Ref.~\cite{wang}}
{$^{\rm c}$ Ref.~\cite{tomanek}}
{$^{\rm d}$ Ref.~\cite{nec60}}

\end{table}

\end{document}